\documentclass[aps,pre,twocolumn,superscriptaddress,amsmath,amssymb]{revtex4-2}

\usepackage[utf8]{inputenc}
\usepackage{amsmath,amsthm}
\usepackage{graphicx}
\usepackage[colorlinks = true,
            linkcolor = blue,
            urlcolor  = blue,
            citecolor = blue,
            anchorcolor = blue]{hyperref}
\usepackage{braket}
\usepackage{xcolor}
\newcommand{\tr}[1]{\ensuremath{\operatorname{tr}\!{#1}}}

\newcommand{\bea}{\begin{eqnarray}}
\newcommand{\eea}{\end{eqnarray}}

\newcommand{\bu}{\mathbf{u}}

\global\long\def\ga{\gamma} \global\long\def\de{\delta}
\global\long\def\De{\Delta}

\global\long\def\dL{\mathbb{L}}

\global\long\def\ell#1{\theta_{#1}}

\global\long\def\si{\sigma}

\global\long\def\eps{\epsilon}
\global\long\def\al{\alpha}

\global\long\def\ga{\gamma} \global\long\def\de{\delta}

\global\long\def\no{\nonumber}

\theoremstyle{thm@}

\theoremstyle{remark}

\def\dA{\mathbb{A}}
\def\dL{\mathbb{L}}
\def\dOmega{\mathbf{\Omega}}

\begin{document}

\title{
Exact Steady State of a One-end Driven XXZ Spin Chain with Boundary Field
}
%$XX0$  spin-$\frac12$ chain }

%\author{Xin  Zhang}
%\affiliation{Beijing National Laboratory for Condensed Matter Physics, Institute of Physics, Chinese Academy of Sciences, Beijing 100190, China}
%\author{  Andreas Kl\"umper} 
 %\affiliation{Department of Physics,
 % University of Wuppertal, Gaussstra\ss e 20, 42119 Wuppertal,
 % Germany}

\author{Vladislav Popkov}
\affiliation{Faculty of Mathematics and Physics, University of Ljubljana, Jadranska 19, SI-1000 Ljubljana, Slovenia}
 \affiliation{Department of Physics,
  University of Wuppertal, Gaussstra\ss e 20, 42119 Wuppertal,
  Germany}
\author{Toma\v z Prosen}
\affiliation{Faculty of Mathematics and Physics, University of Ljubljana, Jadranska 19, SI-1000 Ljubljana, Slovenia}
\affiliation{Institute of Mathematics, Physics and Mechanics, Jadranska 19, SI-1000 Ljubljana, Slovenia}

\begin{abstract}
We find an exact nonequilibrium steady state of an open dissipatively driven XXZ spin-$\frac12$ chain with source or sink spin bath at one end and an arbitrary boundary field at the other end.  
%can be expressed via  usual  $SU_q(2)$  Lax representation,  with modified right boundary auxiliary vector
\end{abstract}
\maketitle

One of fascinating discoveries in the field of integrable quantum systems,  exposed to locally acting dissipation at the boundaries is that  many properties of their nonequilibrium steady
state (NESS),   including the complete NESS density operator itself, can be expressed in Matrix Product Ansatz (MPA) form with an infinitely dimensional auxiliary space \cite{2011Prosen}.  This MPA construction led to  further discoveries,  even in the field of fully coherent quantum systems (i.e. without dissipation), 
e.g.  of quasi-local charges  \cite{2013ProsenIlievski,2015ProsenReview}. These complemented  the set of strictly local symmetries,  previously known from the Quantum Inverse Scattering Method~\cite{Korepin}, and were instrumental in explaining high-temperature spin ballistic transport~\cite{2013ProsenIlievski}. The MPA-NESS formalism was originally formulated for a one-dimensional  setup which describes a coherent evolution in the bulk 
and dissipative coupling at both boundaries  \cite{2015ProsenReview}.   However the MPA  appears to have a rather straightforward generalization to a mixed,  or hybrid, case, 
where  the dissipation acts at one boundary only,  and at the other boundary it is substituted with a coherent external field term.  

The goal of our short communication is a simple algebraic derivation of  MPA of NESS for a hybrid scenario,  where the underlying coherent model is a 
paradigmatic XXZ 
spin -$\frac12$ Heisenberg Hamiltonian,  see below.  Our method is straightforwardly generalizable to other Yang-Baxter integrable spin chains with hybrid driving. 
%We note that an example of this kind (for a specially chosen driving field) 
% was  treated in a recent paper \cite{2025Clerk}.

 %An example of this kind
%was considered in recent paper \cite{2025Clerk}, where it is  treated via a
%rather technical  “isolated defect operator” method of of \cite{2011Prosen},  resting  upon  many qubic algebraic relations
%for the Lax operator components,   making the results   rather obscure.  As a result,  an origin of the fundamental recurrence relation  (S.14) in the Supplement o
 %is difficult to trace,  without resolving to computer-assisted symbolic simulations.  

We aim at finding  a unique  non-equilibrium steady state (NESS) of a quantum spin chain with a dissipator at left end 
and and arbitrary coherent driving (a boundary field) at the other end,  described by Lindblad Master equation 
\begin{align}
&\frac{\partial \rho}{\partial t}= -i [H,\rho] + \ga {\cal D }_{\si_1^+}[\rho]= {\cal L}[\rho] , \label{LME}\\
&H = \frac12 
\sum_{n=1}^{N-1}\left(
 \si_n^x   \si_{n+1}^x + \si_n^y   \si_{n+1}^y + \De \si_n^x   \si_{n+1}^x  
\right)+g_N,\no \\
 &{\cal D }_{F}[\rho]= 2 F \rho F^\dagger - \left\{\rho, F^\dagger F \right\} .\no
\end{align}
  Here $g_N = \vec{g}\cdot\vec{\sigma}_N$ is an arbitrary field,  locally acting at the rightmost spin $N$ of a chain.
Similar scenario,  for a special choice  $\vec{g} = (g,0,0)$ was considered in recent paper \cite{2025Clerk}  and treated via a
%Their central result,  the exact solution Eq.(4),  contains the right boundary vector,  coefficients of which are  given by  a recurrence (S.14) in %Supplemental.  The recurrence (S.14) is produced by 
a quite technical
 “isolated defect operator” method of \cite{2011Prosen},  resting  upon validity of qubic algebraic relations
for the Lax operator components.  As a result,  the origin of some  main results of \cite{2025Clerk},  e.g.  of a fundamental 
recurrence  (S.14) in the Supplemental
of  \cite{2025Clerk},
 is difficult to trace,  without resolving to  symbolic computer-assisted calculus.  Within our algebraic approach,  the same recurrence,  and, in addition,  for 
completely general field $\vec{g}$, appears as a  consequence of the fundamental  relations for 
$U_q(SU(2))$ quantum group generators
$S_z,S_\pm$.

 Lax operator of the XXZ spin-chain with dissipation   \cite{2013MPA,2015ProsenReview},  in terms of  
 $SU_q(2)$ generators $S_z,S_\pm$, reads  
\begin{align}
&L_n= \sum_{\al=x,y,z} L^\al \si_n^\al = 
\left(
\begin{array}{cc}
[S_z]_q & S_{+} \no \\
S_{-} & -[S_z]_q
\end{array}
\right),  \no\\
&[S_z]_q = \sum_{k=0}^\infty [s-k]_q \ket{k} \bra{k},\no \\
&S_{+} =  \sum_{k=0}^\infty [k+1]_q \ket{k} \bra{k+1}, \label{eq:L} \\
&S_{-} =\sum_{k=0}^\infty [2s-k]_q \ket{k+1} \bra{k}\no,\\
&[x]_q = \frac{q^x -q^{-x}}{q-q^{-1}},\no
\end{align}
where $s\in\mathbb C$ is a representation parameter (will be specified later), and $q$ is related to the XXZ anisotropy parameter 
$\De$ by
\begin{align}
 \De = \frac {q + q^{-1}}{2},\label{eq:De}
\end{align}
and is real or unimodular in the easy-axis $|\Delta|> 1$ or easy-plane $|\Delta|<1$, regime, respectively.
$S_z,S_\pm$  satisfy canonical 
commutation relations of $U_q(SU(2))$, while  
$\bra{0}$ is the lowest weight vector of the  irrep $\bra{0} S_{-}  = 0$.

It follows that $L_n$ satisfies a
divergence relation \cite{2013MPA}
\begin{align}
&\left[\frac12 h_{n,n+1}, L_n L_{n+1} \right] = A_n L_{n+1} - L_{n} A_{n+1}, \label{divergence}\\
&A_n  =  \frac{q^{S_z} +q^{-S_z}}{2} 
\left(
\begin{array}{cc}
1 & 0\\
0& 1
\end{array}
\right),
\end{align}
where $h_{n,n+1} = \si_n^x   \si_{n+1}^x + \si_n^y   \si_{n+1}^y + \frac{q + q^{-1}}{2} \si_n^x   \si_{n+1}^x   $ is 
the density of the XXZ Hamiltonian with the uniaxial anisotropy $\De$, see (\ref{eq:De}).
 The unique time-independent solution of Eq.~(\ref{LME}), ${\cal L} [\rho_{\rm NESS}]=0$,
 the so-called  dark state,  or  NESS, 
 is postulated as \cite{2011Prosen}
\begin{align}
&\rho_{\rm NESS} \sim \dOmega_N \  \dOmega_N^\dagger,  \ \dOmega_N =\bra{\bu_{\rm l}} L_1 L_2  \ldots L_N \ket{\bu_{\rm r}}.
\label{eq:NESS}
\end{align}
 Writing $\dOmega_N^\dagger =\bra{\bu_{\rm l}^*} M_1 M_2  \ldots M_N \ket{\bu_{\rm r}^*}$,
where $M_n = \sum_\al M^\al \si_n^\al$,
$M^\al=(L^\al)^*$, we rewrite the above as 
\begin{align}
&\rho_{\rm NESS} \sim \bra {\bu_{\rm l}} \otimes  \bra {\bu_{\rm l}^*}  \dL_1  \ldots \dL_N  \ket {\bu_{\rm r}} \otimes  \ket {\bu_{\rm r}^*}, \label{rhoNESS} \\
&\dL_n = L_n \underset{a}{\otimes} M_n \no
\end{align}
where $\underset{a}{\otimes}$ denotes the tensor product in the auxiliary space while the usual matrix product is taken in physical space. 

From (\ref{divergence}) we obtain a divergence condition for $M_n$ 
\begin{align}
&[h_{n,n+1}, M_n M_{n+1}] = B_n M_{n+1} - M_{n} B_{n+1}, \no\\
& B_n = -A_n^*, \label{eq:B}
\end{align}
which leads to a divergence condition for $\dL_n$
\begin{align}
&[h_{n,n+1}, \dL_n \dL_{n+1}] = \dA_n \dL_{n+1} - \dL_{n} \dA_{n+1},\no\\
&\dA_n = A_n M_n+L_n B_n.  \label{eq:dA}
\end{align}
and consequently to
\begin{align}
&\left[\sum_{n=1}^{N-1} h_{n,n+1} + g_N,   \dL_1  \ldots \dL_N \right] = \no \\
&\dA_1 \otimes \dL^{\otimes_{N-1}} - \dL^{\otimes_{N-1}} \otimes \left(\dA_N -
 [g_N,\dL_N ] \right). \no
\end{align}

The dark state condition ${\cal L}[ \rho_{\rm NESS}]=0$  for open XXZ Hamiltonian with arbitrary right boundary field $g_N$
is satisfied via imposing two local conditions,  for the left and for the right boundary:
\begin{align}
 &\bra {\bu_{\rm l}} \otimes  \bra {\bu_{\rm l}^*} (-i \dA_1 + \ga {\cal D}_{\si_1^{+}} [\dL_1])=0, \label{LB} \\
&\left( \dA_N - [g_N,\dL_N ] \right) \ket {\bu_{\rm r}} \otimes  \ket {\bu_{\rm r}^*} =0. \label{RB}
\end{align}
 Eq.~(\ref{LB}) is satisfied by setting  $\bra {\bu_{\rm l}}= \bra {0}$ (the lowest weight state in auxiliary space),
and an appropriate choice of  the  representation parameter $s$ in (\ref{eq:L})  tuned to dissipation strength $\ga$ via 
\begin{align}
&\ga= i \frac{q^s +q^{-s}}{2 [s]_q},  \label{eq:s(gamma)}
\end{align}
see \cite{2013MPA,2015ProsenReview}.  For the isotropic limit $\De=1$, or $q=0$,  Eq.~(\ref{eq:s(gamma)}) reduces to: $\ga = i/s$. 
For  treating Eq.~(\ref{RB}) we first calculate,  summing over repeated indices $a,b,c,d,e$:
\begin{align*}
 & [g,\dL ]= [g_a \si^a, L^b \si^b M^c \si^c] = g_a  L^b  M^c [\si^a, i \eps_{bcd} \si^d]  \\
&= -2 g_a  L^b  M^c  \eps_{bcd} \eps_{ad e} \si^e= 2 g_a  L^b  M^c (\de_{ab} \de_{ce} - \de_{be} \de_{ca}) \si^e\\
& = 2 (\vec {g}\cdot \vec{L}) T- 2 L (\vec {g}\cdot \vec{T}),
\end{align*}
(a crucial point in the calculation is  $L_n$ being  traceless in the physical space).
Accounting for (\ref{eq:B}), (\ref{eq:dA}), and using Hermiticity of the field term $g_N$,  
 Eq.~(\ref{RB}) reduces to 
\begin{align}
\left(A - 2 \sum_{\al=x,y,z} g_\al L^\al \right)\ket {\bu_{\rm r}} =0, \label{resHRcomm}
\end{align}
which is a three-point recurrence relation for the coefficients of $\ket{\bu_{\rm r}} = \sum_{j=0}^\infty u_j \ket{j}$. 
However, since $\bra{\bu_{\rm l}}=\bra{0}$,  only the coeffients $u_j$ with $j\leq N$ give nonzero 
contribution to NESS (\ref{eq:NESS}).
Explicitly,  denoting 
$g_{\pm} = g_x \pm i g_y$,  
$a_j=\frac12 ( q^{s-j} + q^{j-s})$,   the recurrence relation (\ref{resHRcomm}) for the components 
reads
\begin{align}
&\left(-\frac{a_j}{2} + [s-j]_q \ g_z \right) u_j + g_{-}[2s-j+1]_q  u_{j-1} \no \\
&+ g_{+} [j+1]_q  u_{j+1}=0, \quad j=0,1,2, \ldots 
 \label{eq:recurr}\\
&u_{-1}=0,\ u_0=1. \no
\end{align}
The condition  $u_{-1}=0$ is needed for a formal validity of (\ref{eq:recurr})  for $j=0$,  and  
  $u_0=1$ can be chosen without losing generality. 
%Alternatively,  using definitions (\ref{eq:dcoeff}),   (\ref{eq:recurr}) can be written as 
%\begin{align}
%&\left(-\frac{a_j}{2} + d_j \ g_z \right) u_j + g_{-} d_{j-1}^{-}  u_{j-1} + g_{+} d_j^{+}  u_{j+1}=0, 
%\label{eq:recurr1}
%\end{align}

The  recurrence (\ref{eq:recurr}) is always well defined,  except for zero field $\vec{g}=0$ or the field along $z$-axis
$\vec{g}=(g_z,0,0)$. In the latter 
 case
the assumption $u_0=1$ becomes invalid,  and
the  $\bu_{\rm r}$ components should be chosen as $u_j=\de_{j,N}$:
the corresponding $\dOmega_N \sim (\si^+)^{\otimes N}$ and the 
unique solution  of (\ref{LME}) is a vacuum 
 pure state $\rho_{\rm NESS}= 
\ket{\uparrow \uparrow \cdots \uparrow }\bra{\uparrow \uparrow \cdots \uparrow }$.

We conclude summarizing  our main result:  

The unique NESS of (\ref{LME}),   $ {\cal L}[\rho_{\rm NESS}]=0$ 
with $g_N=\frac12 (g_{+}\si_N^{-} + g_{-}\si_N^{+})+ g_{z}\si_N^{z}$,  is given by 
\begin{align}
&\rho_{\rm NESS} = \frac {\dOmega_N \,  \dOmega_N^\dagger}{\tr( \dOmega_N\,\dOmega_N^\dagger)},    \no\\
 & \dOmega_N =\bra{0} L_1 L_2  \ldots L_N \ket{\bu_{\rm r}}, \label{eq:NESSfinal} \\
&\ket{\bu_{\rm r}}= \sum_{j=0}^\infty u_j \ket{j} \no
\end{align}
with $L_n$ from (\ref{eq:L}),  $s$ from (\ref{eq:s(gamma)}),  $q$ from (\ref{eq:De}),  and 
 coefficients $u_l$ given by the linear recurrence (\ref{eq:recurr}).

\textit{Remark 1.~~}
The NESS of a modified Liouvillian (\ref{LME}) with boundary field $g_N$ removed and another boundary dissipator  
  added    
${\cal L}[\rho] \rightarrow {\cal L}[\rho] +\ga {\mathcal{D}}_{\si_N^-}[\rho] $
 is given by the same expression (\ref{eq:NESSfinal}) with
$\bu_{\rm r} = \ket{0}$. 

\textit{Remark 2.~~}
Replacing the left dissipator ${\cal D}_{\si_1^{+}}$ in (\ref{LME}) by ${\cal D}_{\si_1^{-}}$ one should transpose the Lax operator in (\ref{eq:L}). All the calculations are then performed in full analogy. 
In that case,  and for a special choice of the boundary field $\{g_x,g_y,g_z\} = \{ \Omega/2, 0,0\}$,  one recovers 
the Eq. (S.14) from the Supplemental material of \cite{2025Clerk} 
Eq.(\ref{eq:NESS}) is equivalent to the Eq.(4) of \cite{2025Clerk}.

\begin{acknowledgements}
  V.P. and T.P.  acknowledge support by ERC Advanced grant
  No.~101096208 -- QUEST, and Research Program P1-0402 and Grant N1-0368 of Slovenian
  Research and Innovation Agency (ARIS). V.P. is also supported by
  Deutsche Forschungsgemeinschaft through DFG project
  KL645/20-2. 
\end{acknowledgements}

\end{document}